Correspondence and requests for materials should be addressed to L.S.B. (bouchard@chem.ucla.edu)

# Nanodiamond Landmarks for Subcellular Multimodal Optical and Electron Imaging

Mark A. Zurbuchen[1,2,3], Michael P. Lake[4], Sirus A. Kohan[5], Belinda Leung[4], Louis-S. Bouchard[2,4,6,7]

[1]Department of Materials Science and Engineering, [2]California NanoSystems Institute, [3]Western Institute of Nanoelectronics, Department of Electrical Engineering. [4]Department of Chemistry and Biochemistry, [5]The Electron Microscopy Services Center of the Brain Research Institute, [6]Department of Bioengineering; University of California, Los Angeles, CA 90095, USA. [7]Jonsson Comprehensive Cancer Center (UCLA).

**There is a growing need for biolabels that can be used in both optical and electron microscopies, are non-cytotoxic, and do not photobleach. Such biolabels could enable targeted nanoscale imaging of sub-cellular structures, and help to establish correlations between conjugation-delivered biomolecules and function. Here we demonstrate a sub-cellular multi-modal imaging methodology that enables localization of inert particulate probes, consisting of nanodiamonds having fluorescent nitrogen-vacancy centers. These are functionalized to target specific structures, and are observable by both optical and electron microscopies. Nanodiamonds targeted to the nuclear pore complex are rapidly localized in electron-microscopy diffraction mode to enable "zooming-in" to regions of interest for detailed structural investigations. Optical microscopies reveal nanodiamonds for *in-vitro* tracking or uptake-confirmation. The approach is general, works down to the single nanodiamond level, and can leverage the unique capabilities of nanodiamonds, such as biocompatibility, sensitive magnetometry, and gene and drug delivery.**

Nanoparticles have emerged in recent years as a promising approach to particulate labeling probes for multimodal imaging, and also for targeted drug or gene delivery. They can also be used for tracking at the single-molecule level.[1] In general, little is known about the immediate environment of targeted nanoparticles due to a lack of suitable visualization protocols. Knowledge of the local environment would enable clear delineation of relationships between the activity of a labeled agent and the local biological environment. One would like to know, for example, the nature and proximity to adjacent macromolecular assemblies. Despite considerable investigation, the identification and optimization of such a cross-platform biomarker has remained elusive.

Optical microscopy offers versatility, specificity, and sensitivity in fixed and live cell settings.[2] The development of super-resolution techniques has improved spatial resolution to tens of nanometers, but these techniques are restricted to a subset of cellular processes.[3-5] Transmission electron microscopy (TEM) offers higher spatial resolution.[6] But, it lacks reliable multimodal markers to align the field-of-view with a desired subcellular region. This is particularly true if multiple iterations between imaging techniques is desired. Routine multimodal correlated imaging, which could enable the study of live cells with concurrent visualization of ultra-structural details,[7] would require the development of inert biomarkers and protocols for correlating optical and electron microscopies.

Reporter systems suitable for both optical and TEM imaging are lacking. For example, fluorescent dyes and fluorophores, required for fluorescence microscopy, cannot be resolved by TEM, can be cytotoxic, and photobleach under optical excitation. Several candidate markers for multimodal imaging have been pursued, but none has yet been found to be practical and universally applicable. Gold nanoparticles and nano-gold cluster compounds are readily bio-conjugated and resolvable by TEM due to their strong electron scattering. Gold can be cytotoxic by itself, and frequently becomes isolated in lysosomes in live cells. Detergent can be used after fixation to allow penetration into cells, but penetration and immunolabeling efficiency can be low, particularly with colloidal gold, due to steric hindrance.[8] Particulate markers such as gold particles are frequently applied after cryo-slicing and thawing, after cryo-fracture, or post-embedding after slicing. That is, after cell death. These approaches have higher immunolabeling success rates, which is dependent upon particle size and whether or not a slice is embedded.[8] While the cryo-slicing approach has produced some impressive correlative imaging results,[9] it is an exceptionally difficult approach, and is also not amenable to multiple iterations between imaging techniques. Last, immunogold without fluorophores can be used as a particulate marker, but for typical particle sizes, requires silver-enhancement to enlarge the particles to render them visible in optical and/or electron microscopies of slices.[10]

Colloidal quantum dot (QD) biolabels are resolvable by TEM, do not photobleach, and can be functionalized with antigen-specific antibodies for targeting. They are, however, cytotoxic, so their surfaces must be decorated with organic conjugants to provide a barrier.[11] They (QDs) also suffer from "blinking" problems[12] and are incompatible with the osmium tetroxide stain frequently used in sample preparation for TEM.[11,13]

Fluorescent nanodiamonds (NDs) have emerged as promising biomarkers. They have been used as optical labels,[14] magnetic sensors,[15] magnetic resonance imaging (MRI) contrast agents,[16] cell division monitors,[17] drug and gene delivery vectors,[18,19,20] and for single-molecule tracking in live cells,[21] and for nanometer-scale thermometry in living cells.[22] Fluorescence can be enhanced by implanting strongly fluorescing nitrogen vacancy ($NV^-$) color centers[23] to enable optical imaging. Such diamond labels are amenable to transfection into the cytoplasm,[33] have surfaces that can be conjugated directly to proteins for targeting sub-cellular structures, are non-toxic to cells[24,25] and microorganisms,[26] do not photobleach or blink, and are compatible with traditional TEM stains and labels.

In this Letter, we describe the use of bioconjugated NDs[6,27] as markers for locating target cell structures. Locating ND landmarks by regular TEM is problematic due to the strong background signal originating from the amorphous fixation medium. Nanodiamonds tend to agglomerate,[28,29] reducing the bio-availability of conjugants, which tend to remain outside of cells, adhere to the outer cell membrane,[30] and when taken up generally remain trapped in endosomes,[30-32] all of which render them biologically inert. While the successful transfection of conjugated NDs has been reported using fluorescence microscopy, the results have not been confirmed independently.[33] To this end, lattice-resolution TEM or electron diffraction is required.

Our NDs are implanted with $NV^-$ centers and targeted using a nuclear membrane-specific localizer protocol where single NDs are transfected into the cytoplasm of live HeLa cells. After embedding and slicing, the cells are imaged by fluorescence lifetime imaging (FLIM), confocal microscopy and by TEM — bright-field, lattice-fringe imaging, and electron diffraction — to unambiguously differentiate NDs from similar-looking features. The bright fluorescence of NDs yields excellent cellular labels. Confocal fluorescence microscopy confirms that control (untransfected) cells incubated with an excess of NDs primarily form aggregates of NDs outside cells, with occasional NDs internalized and retained near cell surfaces (Section A, Supplementary Information), consistent with reported observations.[28-32]

**Results**

For lower ND-concentration imaging, we turned to FLIM. DAPI serves as a positive control for lifetime imaging (Fig. 1a), where the nuclei are clearly resolved from the cytoplasm and from areas outside the cell by differential fluorescence lifetimes. The secondary peak in fluorescence is consistent with DAPI having a stronger signal and a clear peak at the expected lifetime of 2.2 ns. This peak can also be resolved into components below 1 ns and above 3 ns, representing bound and unbound DAPI, consistent with the expected emission of DAPI from two and three-photon absorption processes which both occur at 910 nm.[34] This signal dominates. Cellular auto-fluorescence, which had the strongest measured lifetimes between 1 and 3 ns, is consistent with literature values (Fig. 1b). Together these signals serve as strong positive controls for the accurate measurement of fluorescence lifetimes. Nanodiamonds exhibit an extremely short optical emission lifetime, on the order of 200 ps (Fig. 1c). The histograms of lifetimes weighted by pixel intensity show two peaks, one that corresponds to the short emission lifetime of the NDs (~250 ps), and a second that corresponds to the cellular auto-fluorescence and/or DAPI signals. The NDs in cells show the expected distribution, appearing as punctate spots with exclusion from the nucleus (Fig. 1c).

Conjugation and transfection were used to deliver NDs into living cells (*in vitro*), to help them escape from endosomes, and to be released into the cytoplasm. Fluorescence lifetime imaging of untransfected NDs (Figs. 2a,b) confirms the fluorescence imaging results above. Lifetime imaging of cells transfected with NDs conjugated to anti-actin antibodies using polypropylene imide (PPI) dendrimers (Fig. 2c) confirms successful transfection by the broad distribution of NDs throughout the cytoplasm, which is distinct from distributions of nanodiamonds when confined within endosomes or targeted to specific membrane-bound structures. The exclusion of signal from the nucleus is consistent with the published literature on NDs, and provides additional evidence that the NDs are responsible for the short lifetime component. Conjugation alone does not produce a broad cellular distribution, but PPI dendrimers significantly enhance the release of NDs into the cytoplasm (Fig. 2d), promoting conjugation to membrane antibodies.

Imaging of individual NDs in a biological TEM specimen presents several challenges. Because our ultimate goals are to enable multiple iterations of imaging, and ultimately to perform correlated imaging, cell cultures were first embedded in epoxy resin before slicing by ultramicrotome, with slices ranging from 70 to 250 nm in thickness. The mounting resin creates a significant amorphous background that confounds TEM image interpretation, particularly for objects under 100 nm in size. A further complication is that NDs and the mounting resin have essentially identical electron densities. Together, these prevent the use of the most commonly used methods for achieving image contrast in a nano-inclusion. That is, underfocusing the image to achieve Fresnel contrast at edges is not feasible, as contrast delocalization in the mounting medium causes a strongly modulated background that can obscure any such detail, even at very limited defocus. Second, there is no appreciable mass-thickness contrast between NDs (carbon) and the mounting medium (carbon, oxygen, nitrogen). There is also the complicating factor that a stained sample (osmium tetroxide, uranyl acetate and lead citrate) presents significant cell-structure contrast which must be discerned from the ND landmarks.

The TEM technique was initially validated on simulated biological slices, comprising NDs dispersed in agarose gel, fixed, resin-mounted, and sliced for TEM (Section B, Supplementary Information). Our TEM observations of untransfected cells dispersed with NDs reveal the morphology and distribution reported in the literature, and are consistent with our observations made by optical techniques, above. That is, the NDs are loosely agglomerated, and are almost exclusively present in the extra-cellular matrix (Fig. 3a). We found during this study that some dark features having the morphology of a ND particle were in fact not NDs. That is, it is necessary to confirm that any ND-like feature in fact is a ND, rather than, for example, a nano-precipitated lead citrate agglomeration. The second approach to confirm that the dark features are indeed NDs is electron diffraction (Fig. 3b) by the ND lattice. Diffraction spots corresponding to diamond are evident over the diffuse scattering of the amorphous mounting medium. The shadow is due to a pointer used to block the primary beam from the camera, to enable acquisition of diffraction patterns with adequate dynamic range. Without this, camera bloom and oversaturation would obscure the fine details.

**Discussion**

Figure 4 presents the central result of this work. Targeting of the nuclear membrane (Fig. 4a) was accomplished by a covalently conjugated antibody specific to Nup98 (Fig. 4a), which is a nucleoporin (NUP) protein normally found at the nuclear pore, embedded in the nuclear membrane. The particular ND shown in Fig. 4b is composed of three sub-grains. High-resolution transmission electron microscopy (HRTEM) (Fig. 4c) reveals lattice fringes of the crystalline diamond structure, and an electron diffraction pattern of the same ND (Fig. 4d) provides further confirmation that the landmark is a ND. This procedure yielded transfected NDs that were found to be non-agglomerated and attached to the nuclear membrane. The results of Fig. 4 enable us to conclude that TEM can characterize the successful transfection, targeting, and localization (as a landmark or alignment fiducial) of single NDs at a region of interest.

It is anticipated that this work will have a significant transformative impact in the biological sciences, enabling the determination of mechanistic descriptions of biological structures and dynamics. The approach was demonstrated for live-cell labeling, but is expected to be equally powerful for post-slicing labeling, whether by cryo or post-embedding. The intrinsic fluorescence of NDs having $NV^-$ centers means that no fluorophore is needed for optical imaging of the nanoparticulate markers, meaning that cytotoxic fluorophores and detail-obscuring silver enhancement are unneeded. Transfected and targeted NDs could also serve as image-alignment fiducials for tomographic TEM reconstruction. Last, and perhaps most importantly, this $NV^-$ ND approach is amenable to correlated optical and electron microscopy, perhaps even in live cell environments over multiple iterations.

**Methods section**

**Nanodiamonds, Transfection Reagents and Conjugates**

The nanodiamonds used in these experiments were prepared by ball milling of larger ~100 μm diamonds containing $NV^-$ centers and are on average <100 nm in diameter. Conjugation of nanodiamonds to antibodies was performed by oxygen termination of the surface using strong acid treatment followed by EDC conjugation to the antibodies. Polypropylene imide dendrimers were purchased from SymoChem (Holland) and were conjugated with maltotriose to prevent cytotoxicity as described by Mkandawire *et al*.[33]

**Cells and Transfections**

All cells used were HeLa cells grown in DMEM with 10% FBS, 1% P/S. Transfections were performed by mixing transfection reagents with nanodiamond conjugates in HEPES buffer, or nanodiamonds alone in HEPES. Reagents were allowed to precipitate for 20 min at room temperature, followed by drop-wise addition of the solution to cells in serum-free DMEM, 1% P/S for 6 hours. Afterward, the media was replaced with growth media and cells were left to grow for 18-24 hours on plastic coverslips, after which cells were fixed with 2% PFA, and some stained with DAPI. Coverslips were mounted onto slides in 10% PBS / 90% Glycerol and sealed.

**Fluorescence Microscopy**

Confocal microscopy was performed on a Leica SP5-STED microscope. Excitation was done using two laser lines at 514 nm and 548 nm. Samples were bleached by repetitive scanning in order to reduce background. On this system, fluorescence can be detected using a photomultiplier tube (PMT) or avalanche photodiode (APD). Lifetime imaging was performed on a Leica SP2-FLIM microscope with Becker and Hickl SP-830 imaging hardware. Lifetime excitation was performed with infrared light from 900-910 nm with a tunable TI sapphire laser.

**TEM Sample Preparation**

The cell monolayers grown on coverslips were immersed in a solution of 0.1 *M PBS*, pH 7.4 containing 2% glutaraldehyde and 2% paraformaldehyde at room temperature for 2 hours, and then at 4 °C overnight. The cells were subsequently washed in 0.1 *M* PBS buffer and post-fixed in a solution of 1% $OsO_4$ in PBS, pH 7.2–7.4. Samples were then buffered in Na acetate, pH 5.5, and stained in 0.5% uranyl acetate in 0.1 *M* Na acetate buffer, pH 5.5, at 4 °C for 12 hours.

The samples were sequentially dehydrated in graded ethanol (50%, 75%, 95%, 100%) and infiltrated in mixtures of Epon 812 and ethanol (1:1 ratio) and 2:1 for two hours each time. The cells were then incubated in pure Epon 812 overnight and subsequently embedded and cured at 60 °C for 48 hr. Sections of 70-90 nm thickness (gray interference color) were cut on an ultramicrotome (RMC MTX) using a diamond knife. The sections were deposited on carbon-film copper grids and double-stained in aqueous solutions of 8% uranyl acetate at 60 °C for 25 min, and lead citrate at room temperature for 3 min prior to TEM.

**TEM Sample Examination**

TEM examination was performed using an FEI Titan operated at 300 keV. Locations of regions of interest were recorded in relation to a fiducial on the copper grids, enabling location of the same area in subsequent optical microscopy experiments. Low electron beam currents, typically < 0.6 nA, along with standard low-dose imaging techniques, were employed in order to preserve specimen integrity. For very low-magnification imaging, the FEI Titan uses the projector lenses for magnification, the optical path of which blocks many diffracted beams, rendering some NDs visible as dark spots (among real-world artifacts that may also appear dark). At moderate to high magnifications, a restrictive objective aperture is used to block all diffracted beams of the nanodiamond, rendering it dark against a relatively lighter background of the embedded cell. Tilting of an eucentric-height positioned sample was occasionally employed to achieve "twinkling" of ND particles as they moved through strongly diffracting (near-zone axis) orientations. HRTEM images were acquired as close to zero defocus as possible in order to minimize the contribution of the amorphous background, which would otherwise obscure any visible lattice fringes.

Our solution to locating NDs by TEM is (1) to use a very restrictive objective aperture to block all diffracted beams from NDs, making them appear darker than surrounding material, and (2) to exploit their crystalline nature to discern them from sample-preparation artifacts. HRTEM imaging of the lattice fringes of NDs embedded in the amorphous mounting medium is feasible, but only near zero defocus of the objective lens. In the ideal case, lattice fringes in HRTEM images appear strongest at the Scherzer defocus, but due to the limited useful defocus range for cell/ND specimens, this is not possible.

**Acknowledgments**

This study was supported by AFOSR and DARPA QuASAR. We acknowledge the use of instruments at the Electron Imaging Center for NanoMachines (EICN) supported by NIH (1S10RR23057 to ZHZ) at the California NanoSystems Institute (CNSI), UCLA. Confocal laser scanning microscopy was performed at the CNSI Advanced Light Microscopy/Spectroscopy Shared Resource Facility at UCLA. We thank Arek Melkonian for help with preparing solutions. The authors thank Joerg Wrachtrup and Gopi Subramanian for providing the fluorescent nanodiamonds, Hong Zhou, Laurent Bentolila, and Sergey Ryazantsev for helpful discussions, and Charles M. Knobler for critical comments on the manuscript.

**Author contributions**

M.A.Z. developed the TEM methodology and carried out the TEM experiments. M.P.L prepared cultures and performed optical microscopies. B.L. helped with nanodiamond conjugation. S.A.K. devised the agarose control experiment, and prepared TEM samples. M.A.Z, M.P.L and L.S.B. designed experiments. M.A.Z. and M.P.L performed data analysis. M.A.Z., M.P.L., and L.S.B. wrote the paper. M.A.Z. and M.P.L. contributed equally to this work.

**Additional Information**

**Supplementary information:** Supplementary information accompanies this paper at http://www.nature.com/scientificreports

**Competing financial interests:** The authors declare no competing financial interests.

**How to cite this article:**

**Figure captions**

**Figure 1.** Fluorescence lifetime imaging of cells with NDs, with corresponding fluorescence lifetime spectra. (a) HeLa cells stained with DAPI. (b) Unstained HeLa cells. (c) NDs incubated with HeLa cells.

**Figure 2**. Comparison of distributions of nanodiamonds in HeLa cells by fluorescence lifetime imaging. (a) Cellular distribution observed in untransfected NDs (control) in excess concentration. (b) Magnified image of an untransfected cell showing nucleus and cytoplasm, with aggregated nanodiamonds primarily outside of cells. (c) Uniform distributions of NDs throughout the cytoplasm (no DAPI) in NDs transfected into cells with PPI dendrimers. (d) Less uniform distribution of NDs added to cells without transfection agent (brightness enhanced, no DAPI).

**Figure 3**. TEM characterization of a control sample reveals the typical appearance of NDs in non-transfected HeLa cells. (a) Image of several loose agglomerations of NDs in the extracellular matrix. Cytoplasm (Cy), nanodiamonds (ND). Scale bar, 500 nm. (b) Electron diffraction pattern of the same area, confirming the agglomerations to comprise NDs. The shadow is from a pointer, used to block the central beam for image acquisition.

**Figure 4**. TEM verification of ND transfection and adherence to nuclear membrane of HeLa cells. (a) Image of a portion of a cell with a ND adhered to the nuclear membrane. White arrows indicate the nuclear membrane. Stripes are knife-marks from the ultramicrotome, and the wide, horizontal dark band is due to a slight wrinkle of the slice. Scale bar, 500 nm.
(b) Magnified image of single ND (consisting of three domains). Scale Bar, 50 nm.
(c) HRTEM image of the ND, showing lattice fringes (indicated by pairs of lines) which verify its crystalline character. Scale bar, 1 nm. (d) Electron diffraction pattern of the same area, further verifying that the particle is a (crystalline) ND. The shadow is from a pointer, used to block the central beam for image acquisition. Cytoplasm (Cy), Nucleus (Nu), nanodiamond (ND).

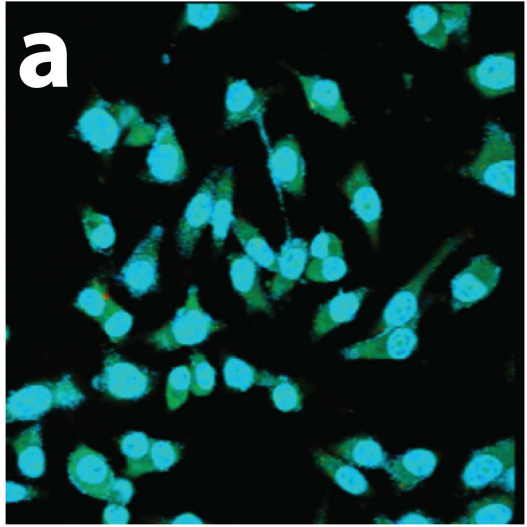 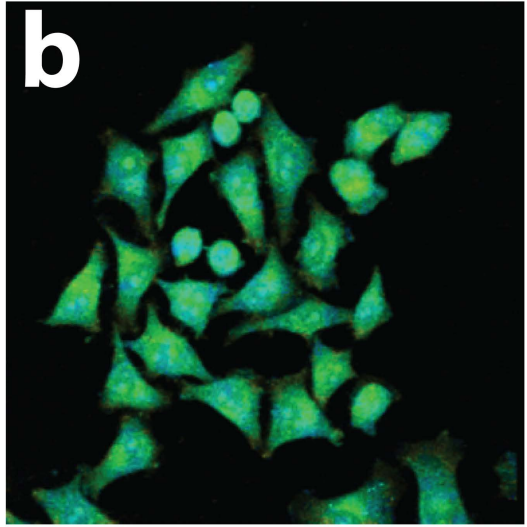 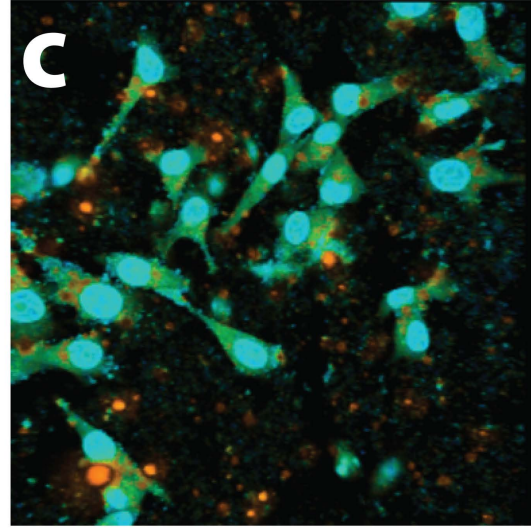
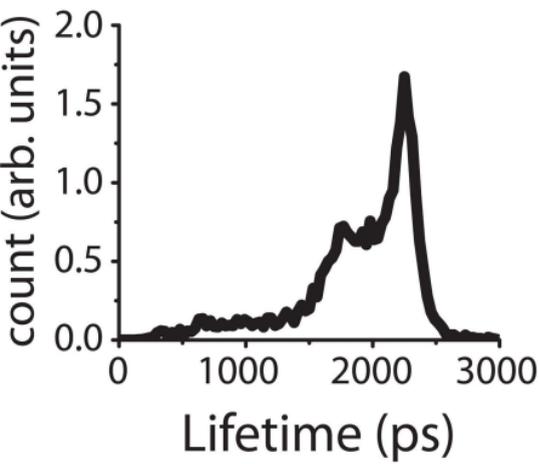 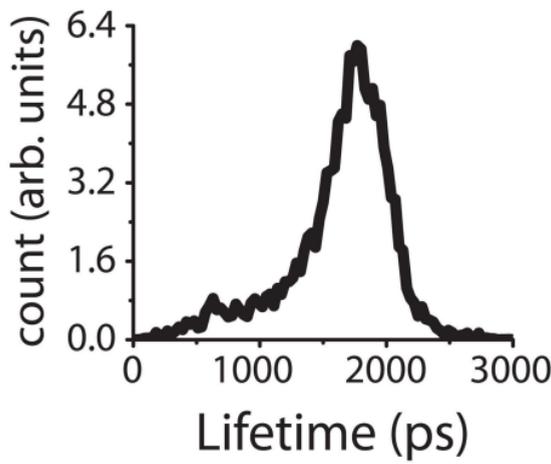 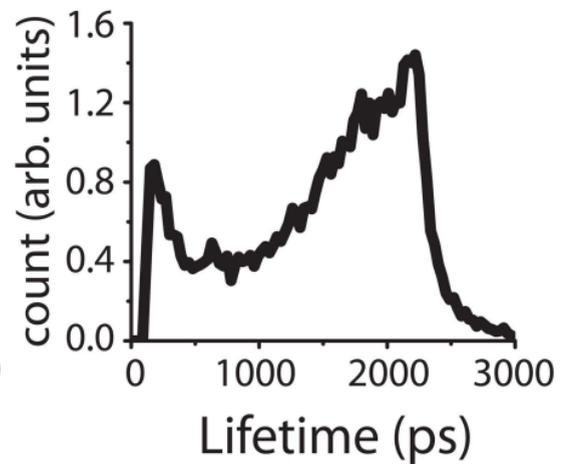

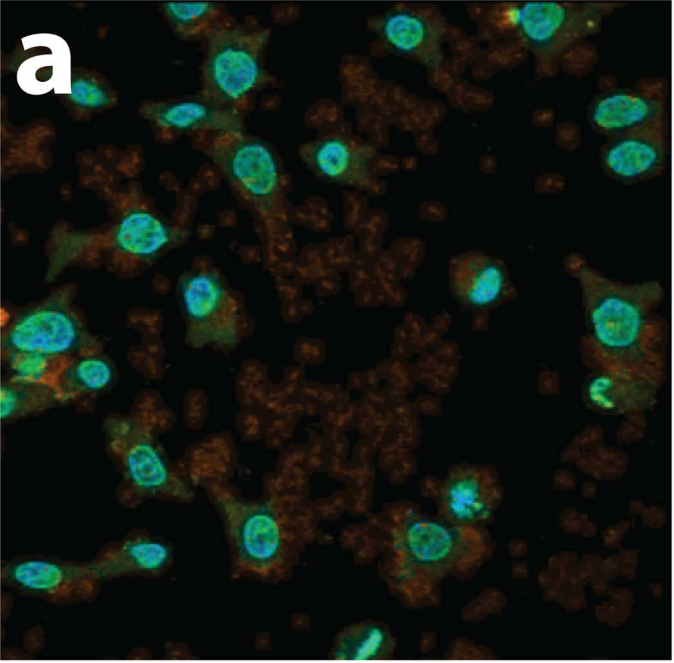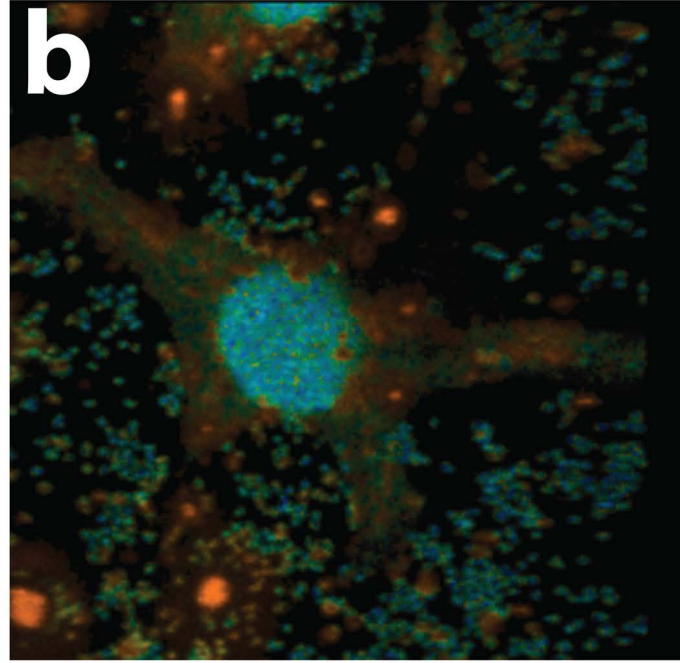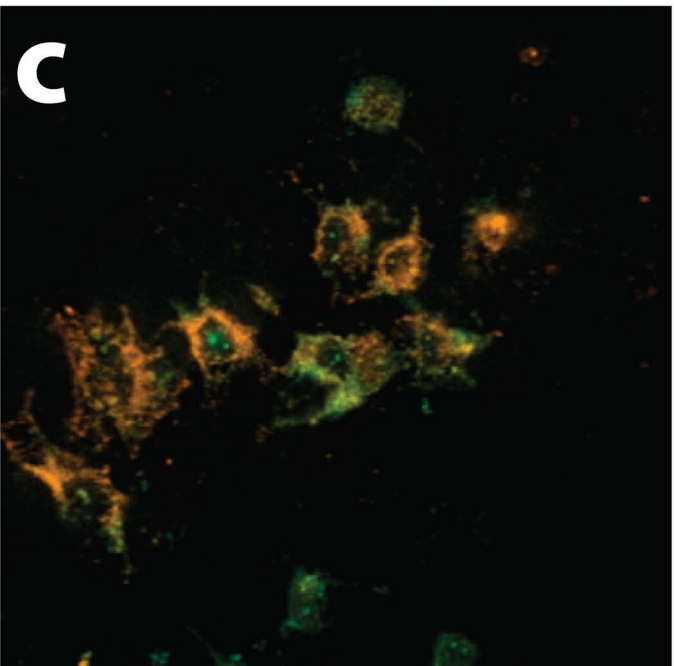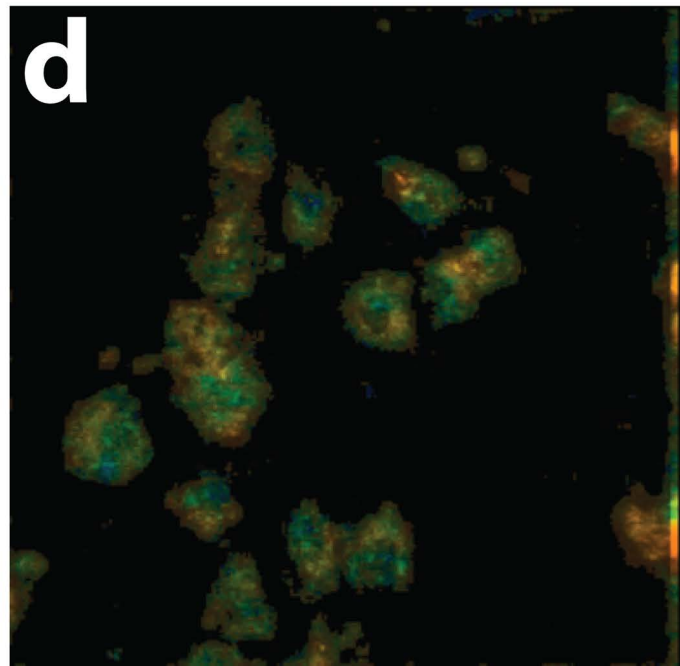

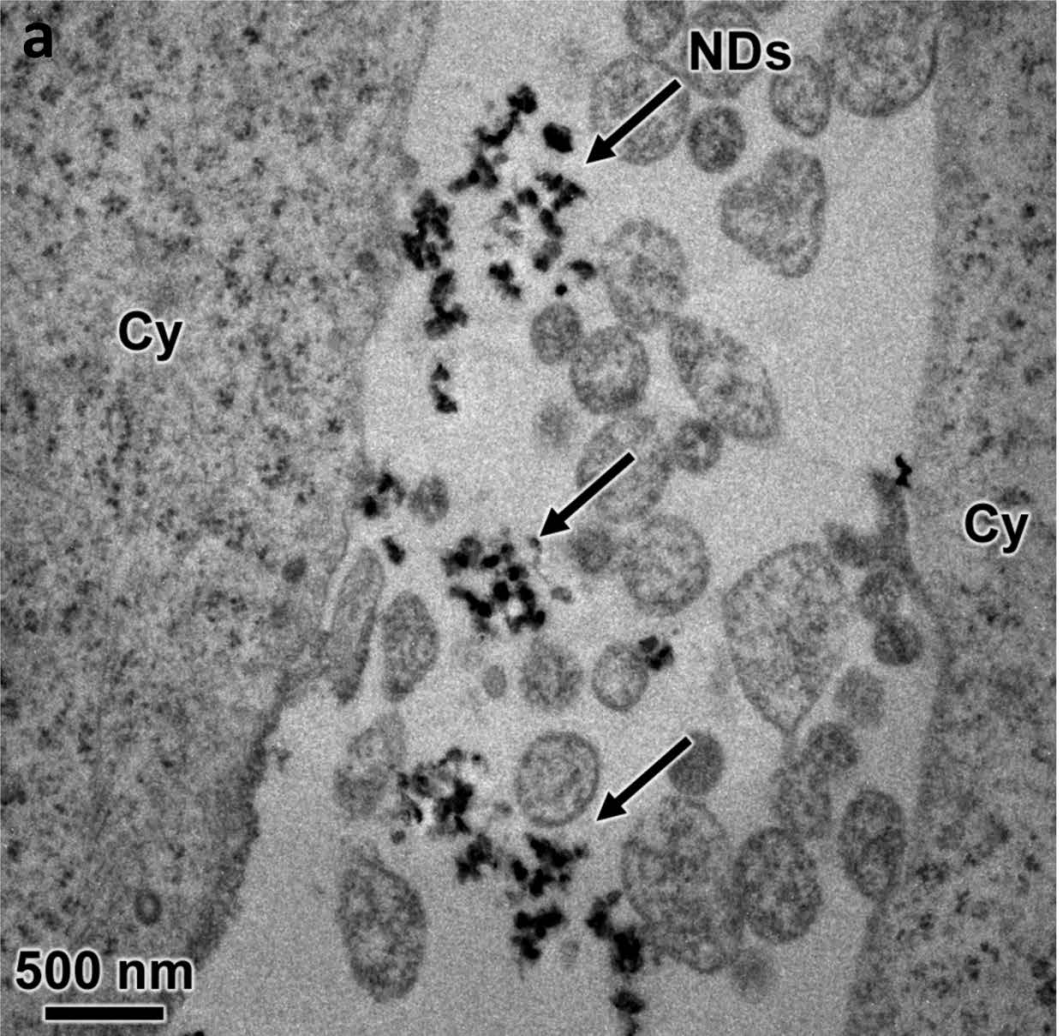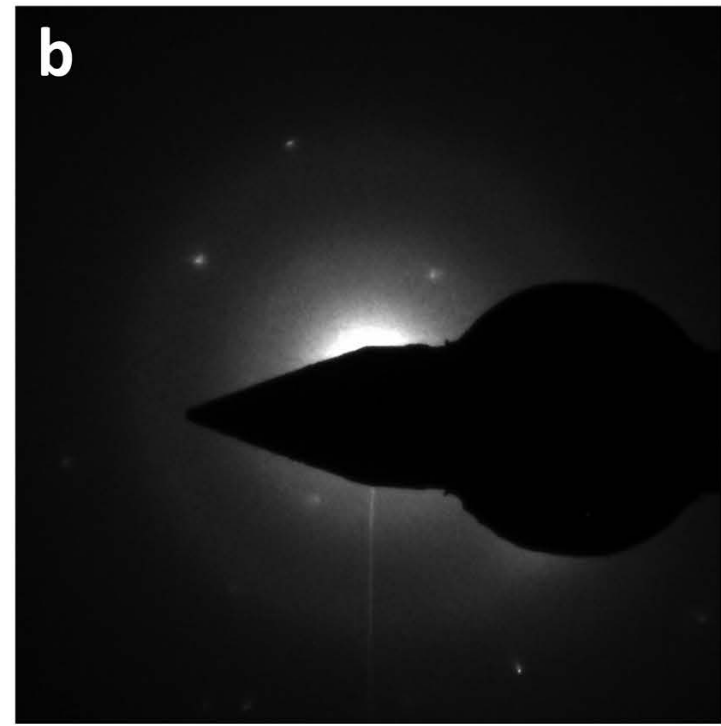

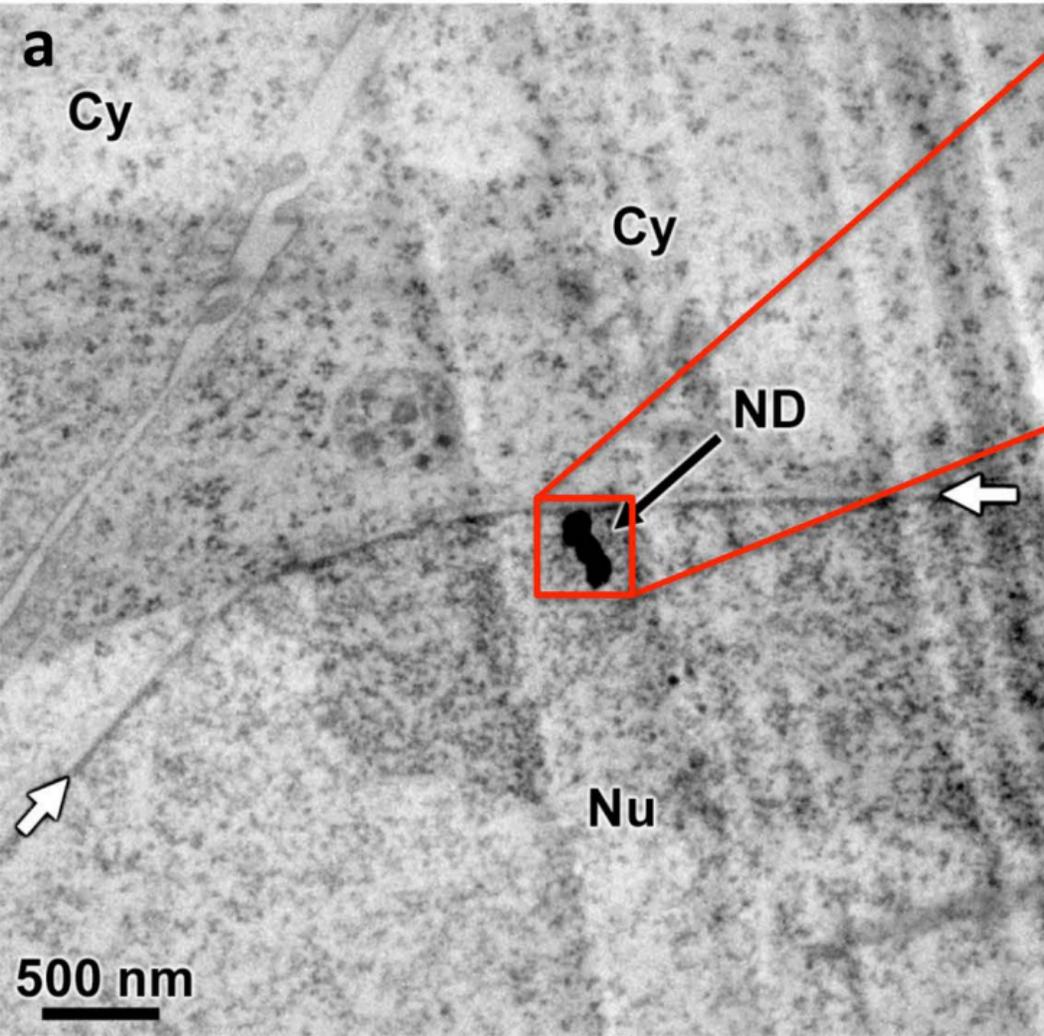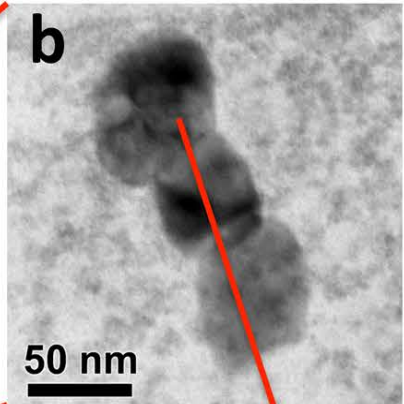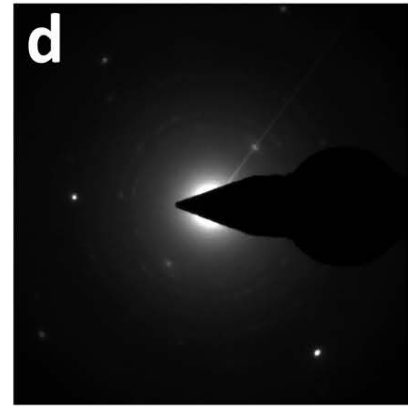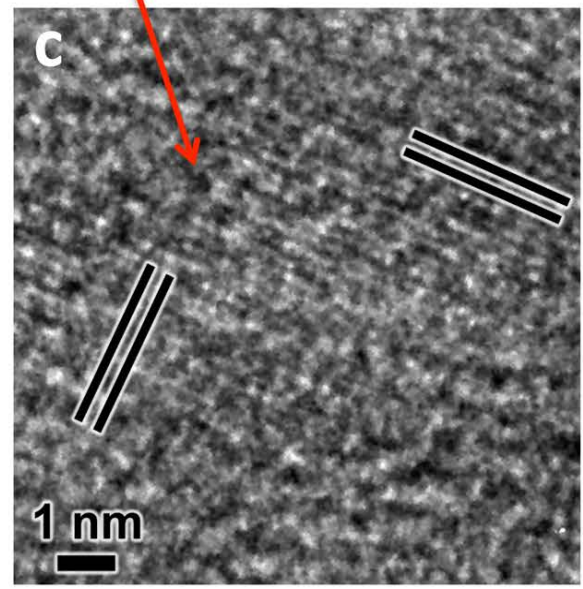